Special Issue on "Complex Organic Molecules (COMs) in Star-Forming Regions"

# Volatile species in comet 67P/Churyumov-Gerasimenko – investigating the link from the ISM to the terrestrial planets

Martin Rubin[1], David V. Bekaert[2], Michael W. Broadley[2], Maria N. Drozdovskaya[3], Susanne F. Wampfler[3]

[1] Physikalisches Institut, University of Bern, Sidlerstrasse 5, 3012 Bern, Switzerland
[2] Centre de Recherches Pétrographiques et Géochimiques, CRPG-CNRS, Université de Lorraine, UMR 7358, 15 rue Notre Dame des Pauvres, BP 20, 54501 Vandoeuvre-lès-Nancy, France
[3] Center for Space and Habitablility, Gesellschaftsstrasse 6, 3012 Bern, Switzerland

## Abstract (150-250 words)

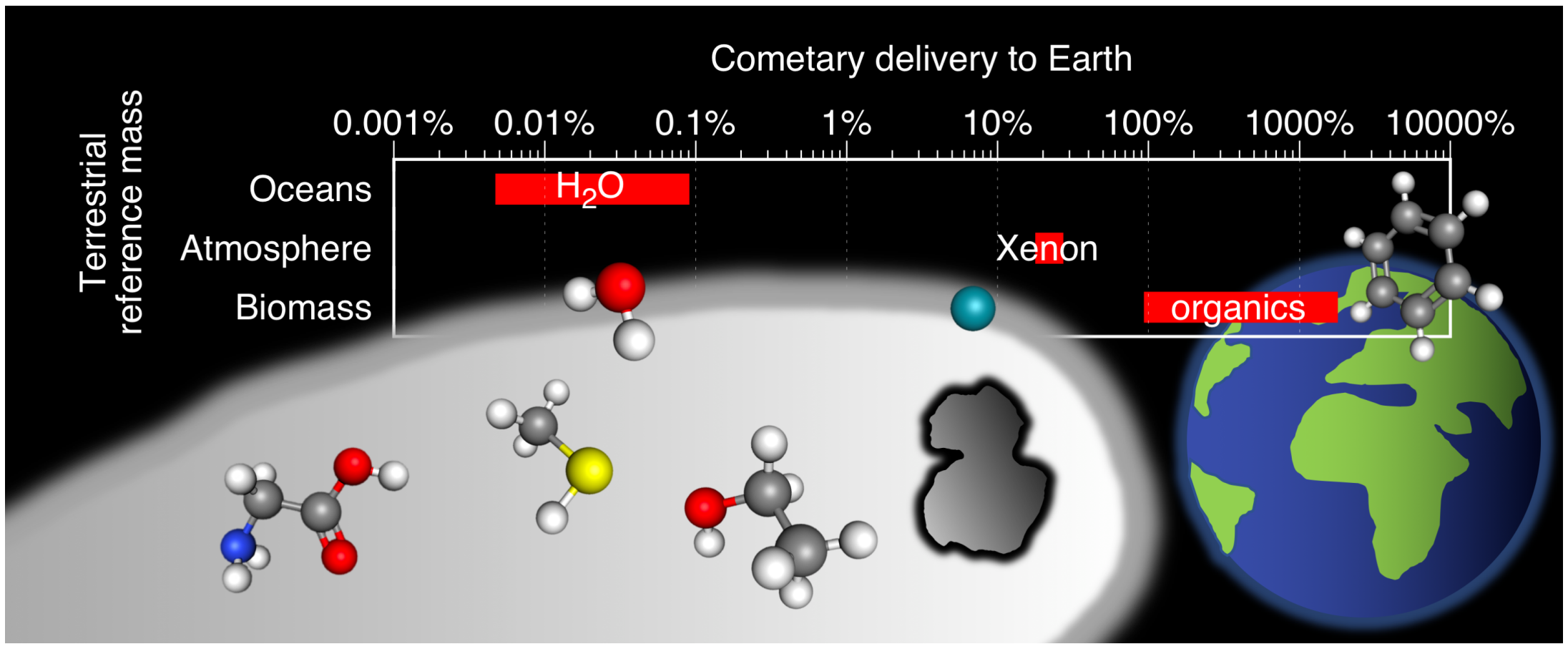

Comets contain abundant amounts of organic and inorganic species. Many of the volatile molecules in comets have also been observed in the interstellar medium and some of them even with similar relative abundances, indicating formation under similar conditions or even sharing a common chemical pathway. There is a growing amount of evidence that suggests comets inherit and preserve substantial fractions of materials inherited from previous evolutionary phases, potentially indicating that commonplace processes occurred throughout comet-forming regions. Through impacts, part of this material has also been transported to the inner planetary system, including the terrestrial planets. While comets have been ruled out as a major contributor to terrestrial ocean water, substantial delivery of volatile species to the Earth's atmosphere, and as a consequence also organic molecules to its biomass, appears more likely. Comets contain many species of pre-biotic relevance and molecules that are related to biological processes on Earth, and have hence been proposed as potential indicators for the presence of biological processes in the search of extraterrestrial life. While the delivery of cometary material to Earth may have played a crucial role in the emergence of life, the presence of such alleged biosignature molecules in the abiotical environment of comets complicates the detection of life elsewhere in the universe.

## Keywords

comets; organic materials; origins; Interstellar medium; planets

## 1. Introduction

The interstellar medium (ISM) is made up of material residing between the stars of galaxies and encompasses regions very diverse in physical and chemical conditions. It ranges from hot ionized gas, where metals are highly ionized, to warm gas with lower ionization states, to diffuse neutral clouds of atomic species, all the way to the few tens of K cold dark molecular clouds where molecules form [1-2] and remain shielded from dissociating radiation.

Complex organic molecules containing at least six atoms including one or more carbon atoms (COMs) are frequently found in some regions of the ISM, i.e. in dark molecular clouds and in the vicinity of protostars [3]. In the solid phase, the presence of oxygenated compounds such as methanol and the tentatively identified ethanol and formic acid is likely a result of grain-surface hydrogenation reactions at cold (~10K) temperatures prior to any energetic processing from protostellar formation. At luke-warm conditions (~40-60 K), the grain-surface pathways are supplemented with radical-radical association reactions [3], in particular if a source of UV photons is available (such as the cosmic ray-induced UV field or protostellar UV photons). These molecules can consequently be released into the gas phase via non-thermal and thermal desorption processes. At temperatures above ~100 K, gas-phase chemical pathways will start to contribute to the synthesis of such molecules [4-6].

It is generally accepted that our solar system originated from the gravitational collapse of a cold dark prestellar core [7]. The collapsing core materials are partitioned between the forming star (our Sun) and the surrounding protoplanetary disk, also called either the protosolar disk or the Solar Nebula for the case of our Solar System. A significant fraction of the core material is also dispersed via, for example, bipolar outflows. Our Solar Nebula is consequently composed of the prestellar dust, gas and ice upon processing via various physical and chemical mechanisms during the collapse phase and later within the protoplanetary disk itself [8-9]. Planets, moons, and minor Solar System bodies all formed from these initial Solar Nebula constituents via complex accretion processes in conjunction with dynamics [10].

Given the abundant amounts of ices in comets, they must have formed at distances from the early Sun where the major volatiles such as $H_2O$, CO, and $CO_2$ [11] were present in the solid form. In today's solar system, the location where solar flux is sufficiently high to thermally desorb water ice is found somewhere between 3 and 5 au. There are different families of comets, originating from the Oort cloud (OC) and from the Scattered disk/Kuiper belt (KB). However, these populations are assumed to have formed closer to the Sun before being scattered through the migration of the Giant Planets either inward, causing the Late Heavy Bombardment [12], or outward to where they reside today [13].

Still, several concurrent models concerning the origin of cometary ices exist, including the presence of copious amounts of amorphous ices inherited from the prestellar core [14], which possibly further underwent (partial) chemical resetting during transition from amorphous to crystalline ice involving, first, the release and then re-trapping of different volatiles in the voids of crystalline water [15-16]. It is therefore important to study the elemental, molecular, and isotopic ratios of the materials in comets, in comparison to laboratory experiments and other objects inside and outside our solar system, to gain insight into the various processes involved in the formation and evolution of comets.

Comets are known to contain significant amounts of organic molecules. Indication of their survival from the preceding cold dark prestellar core all the way to the incorporation into comets is found in the excesses of D, $^{13}C$, and $^{15}N$ isotopes [17-21] in some of the organics collected from comet 81P/Wild 2 by the Stardust mission [22-23]. Similar results are found in parts of the organic material in meteorites and IDPs suggesting an prestellar/protostellar chemical heritage [24-26]. Bockelée-Morvan, et al. [27] also showed that relative abundances of a suite of volatile molecules detected in comet Hale-Bopp resemble the relative abundances measured around newly forming stars. On the other hand, Stardust samples also revealed copious amounts of high-temperature minerals that must have formed much closer to the Sun before being transported outward by radial mixing to the location of formation of comet Wild 2 [28]. Thus, there must be several processes at work in the formation of comets and their relative importance may differ from one comet to another, possibly as a result of their different dynamical histories.

Due to the low gravity in cometary environments [29], the sublimation of different ices releases volatile species that directly escape the nucleus. Molecules, such as methane and formaldehyde, have been observed in numerous comets originating from both, the OC and the KB; (see, e.g., Bockelée-Morvan, et al. [30]). COMs, such as ethane, methanol, and ethanol, have also been identified [31-32]. There are even comet taxonomies based on the amounts of organic molecules, e.g., the three groups of organic-normal, organic-enriched, and organic-depleted [33] occurring in OC and KB families of comets.

The recently concluded Rosetta mission [34-35] of the European Space Agency (ESA) to comet 67P/Churyumov-Gerasimenko (hereafter, 67P/C-G) has more than doubled the number of molecules known to be present in comets [36]. This includes predominantly organic molecules, such as COMs, that have been detected in the vicinity of the nucleus or on its surface. The Rosetta orbiter and its lander Philae carried a comprehensive set of remote sensing and in situ instruments. Gas composition measurements were carried out by the Rosetta Orbiter Spectrometer for Ion and Neutral Analysis (ROSINA; Balsiger, et al. [37]), the Visible and InfraRed Thermal Imaging Spectrometer (VIRTIS; Coradini, et al. [38]), the Microwave Instrument for the Rosetta Orbiter (MIRO; Gulkis, et al. [39]) and the two mass spectrometers Ptolemy [40] and the Cometary Sampling and Composition experiment (COSAC; Goesmann, et al. [41]) on the lander Philae. The similarity of the inventory and relative abundances of volatiles towards the Solar-like protostar IRAS 16293-2422B and comet 67P/C-G provides further evidence that cometary ices are at least in part inherited from the earliest phases of our Solar System's formation [42]. It is unclear whether this holds for the refractory component which includes high temperature minerals most likely formed close to the Sun, as revealed by dust samples returned from comet Wild 2 by NASA's Stardust mission [28].

Comets have long been postulated to be a major supplier of water to Earth [43]. While this has been subject to debate, given the on average elevated D/H ratios in comets compared to terrestrial water [44], sizeable cometary contributions of volatile species to the terrestrial atmosphere are still required [12]. Given that some of the material incorporated in comets may have found its way to Earth and the inner planets, they may also be an important supplier of organic molecules. Therefore, comets may have played a crucial role in the emergence of life which took place soon after the Late Heavy Bombardment [45].

In the next section, we will review the list of species found in both, the ISM and comets. This includes organic, as well as inorganic, molecules. We will discuss the evidence for a common origin and discuss the abundances of a suite of cometary molecules in the following section. Afterwards, we will briefly overview some of these molecules and their relevance to prebiotic chemistry and provide estimates for the total mass delivered to Earth. Finally, there follows a discussion concerning the main conclusions of this paper.

# 2. Composition of cometary ices

Ices contained in the nuclei of comets are subject to sublimation upon solar heating. The size of a cometary nucleus ranges between a few hundred meters to a few tens of kilometers [46] and the low gravity leads to an unbound cometary atmosphere, called the coma, which can be sampled in situ or observed remotely. Assuming spherical expansion in the most simple case [47-48], the density in the near-nucleus coma drops as a function of $r^{-2}$. While the gas may be collisional near the surface, where particle densities can easily exceed $10^{12}\ cm^{-3}$ for an active comet such as 1P/Halley, the density quickly drops when moving away from the comet's surface and the gas becomes non-collisional.

To date, only a few comets have been visited by spacecrafts. Both the Soviet Vega 2 [49] and the European Space Agency's (ESA) Giotto [50] missions passed comet 1P/Halley in 1986, revealing that comets indeed do possess a nucleus resembling the "dirty snowball" proposed by Whipple [29, 51]. Already by the time the flyby of comet 1P/Halley took place, in-situ investigations had proved to be a powerful tool to analyze the composition of the volatile and refractory dust in the coma [52-55]. Since then, several comets have been visited by spacecrafts, including 26P/Grigg-Skjellerup [56], 19P/Borrelly [57], 9P/Tempel 1 [58], and 103P/Hartley 2 [59]. NASA's Stardust spacecraft even brought back dust coma samples of comet Wild 2 [28] and ESA's Rosetta mission carried out a detailed investigation by following comet 67P/C-G [34-35] for more than two years along its orbit, including the deployment of the lander Philae onto the comet's surface [60] in November 2014.

Many more comets have been investigated from ground- and space-based telescopes. Molecules have been detected in the wavelength range covering X-ray (ionized molecules), ultraviolet to visible (electronically excited molecules), infrared (vibrationally excited molecules), and the (sub-)millimeter (rotationally excited molecules) [30, 61]. Using powerful interferometers, such as the Atacama Large Millimeter Array (ALMA), the number of remotely observed molecules in comets [31] and in the in star-forming regions [62] is steadily increasing.

## 2.1. Inventory of cometary volatiles

To date, numerous volatile molecules have been observed remotely and/or in situ at comets. Here, we refer to volatiles as species that sublimate in the temperature regime characteristic for comet 67P/C-G, which ranges from tens of K in the subsurface [63] to several hundred K when considering hot grains that may carry semi-volatiles into the coma [64]. The most abundant volatile species is water, followed by $CO_2$, CO, and others that may be found above or around the percent level with respect to $H_2O$. We refer to these volatiles as major species. Minor species thus make up the rest. There is no strict distinction between the two as relative abundances vary among comets and observations at different heliocentric distances. Hence, this definition is somewhat arbitrary.

Table 1 lists the suite of detected atomic and molecular volatile parent species (i.e., species that are present in the ices in the interior of the comet and are not formed through reactions in the coma after their release) [36] in Hill notation sorted by the number of atoms. The list contains 72 volatile species with at least a tentative detection. Several of the listed molecules may be present in the form of different isomers that in many cases cannot be distinguished by mass spectrometry. This may further increase the diversity among volatile species. The ability to probe the rotational structure of the isomers of facilities such as ALMA is thus crucial to further distinguish between different possible structural isomers and diastereoisomers in comets.

Aside from the heavy noble gases (Ar, Kr, and Xe) [12, 65-66], sulfur has also been detected in its atomic form [67]. The list does not include the species Na, Si, K, and Ca, sputtered from the surface by solar wind and detected in the coma of comet 67P/C-G [68] as they are bound in mineral form on the nucleus [69]. The detailed and long-term investigation of comet 67P/C-G led to the first-time detection of many species [36]. This includes major species such as molecular oxygen [70], $O_2$, noble gases [65-66, 71], and a suite of long chain hydrocarbons [72], which are difficult to observe remotely either due to the absence of a permanent dipole moment, and/or a very low abundance.

Furthermore, numerous isotopologues of the here-listed molecules have been identified at comets, from which isotopic ratios can be derived. For instance, at 67P/C-G, two distinct D/H ratios have been determined in water from the two ratios $D_2O$/HDO and HDO/$H_2O$ [73]. Deviations in oxygen [74], silicon [75], and xenon [71] isotopic ratios with respect to bulk solar system ratios [76] (cf. Hoppe, et al. [77]) have also been identified at 67P/C-G. Some of these isotopic ratios can be used to link the materials in comets to their origin in prestellar cores and will be discussed later in section 3.1.

Table 1 also contains numerous organic and inorganic volatile species. The following section 2.2 focuses on the organic molecules detected in the (semi-)volatile phase, while the refractory component will be only briefly touched upon in section 2.3.

Table 1: Detected volatile parent species (Hill notation) in the ices of 67P/C-G and other comets including organic (Table 2) and inorganic molecules [36] sorted by number of atoms and then by molecular mass. Relative mass abundances, where available in 67P/C-G, can be found in Table 4. Bold: molecules with at least one isomer detected in the ISM [62] with a subset firmly (f), likely (l), and possibly (p) detected in the solid phase [1, 78-79], *italic*: molecules that may be the result of biological processes on Earth and thus relevant as a biosignature to detect life remotely on Earth and elsewhere [80], * tentative detection.

| Number of atoms | | | | | | | | | | | | |
|---|---|---|---|---|---|---|---|---|---|---|---|---|
| 1 | 2 | 3 | 4 | 5 | 6 | 7 | 8 | 9 | 10 | 11 | 12 | ≥13 |
| S | **HF** | ***$H_2O$ (f)*** | ***$NH_3$ (f)*** | ***$CH_4$ (f)*** | ***$C_2H_4$*** | **$CH_5N$** | *$C_2H_6$* | **$C_2H_6O$** (p) | $C_2H_7N$ | *$C_3H_8$* | $C_3H_8O$ | $C_3H_9N$ |
| Ar | ***CO (f)*** | **CHN** | ***$C_2H_2$*** | **$CH_2O_2$** (p) | **$CH_4O$** (f) | **$C_2H_4O$** (p) | **$C_2H_4O_2$** | **$C_2H_5NO$** | **$C_3H_6O$** | **$C_3H_6O_2$** | ***$C_6H_6$*** | $C_4H_8O$ |
| Kr | ***$N_2$*** | ***$HO_2$** | **$CH_2O$** (l) | ***$CH_3Cl$*** | **$C_2H_3N$** | $CH_4OS$ | | ***$C_2H_6S$*** | **$C_2H_6O_2$** | $C_3H_6OS$ | | *$C_4H_{10}$* |
| Xe | ***NO*** | ***$H_2S$*** | **$H_2O_2$** | **$C_3HN$** | **$CH_3NO$** | $CH_4S_2$ | | | $C_2H_5NO_2$ | | | $C_4H_{10}O$ |
| | ***$O_2$*** | ***$CO_2$ (f)*** | **CHNO** | | **$CH_4S$** | | | | $C_2H_6OS$ | | | *$C_7H_8$* |
| | ***HCl*** | ***COS (l)*** | **$CH_2S$** | | | | | | | | | *$C_7H_6O_2$ |
| | **OP** | ***$O_2S$*** *(p)* | $C_2N_2$ | | | | | | | | | *$C_5H_{12}$* |
| | **OS** | *$CS_2$* | $S_4$ | | | | | | | | | $C_5H_{12}O$ |
| | $S_2$ | $S_3$ | | | | | | | | | | *$*C_8H_{10}$* |
| | HBr | | | | | | | | | | | *$*C_{10}H_8$* |
| | | | | | | | | | | | | *$C_6H_{14}$* |
| | | | | | | | | | | | | *$C_7H_{16}$* |
| | | | | | | | | | | | | *$*C_8H_{18}$* |

## 2.2. Organics in the (semi-)volatile phase

Table 2 lists the subset of organic molecules detected in comets from Table 1. Identification by mass spectrometry is not always unique due to isomerism where the same molecular formula exhibits different structural and spatial arrangements. Hence, several molecules listed in Table 2 may also be present in the form of a different compound. One possible way to distinguish among the different isomers is by their characteristic fragmentation patterns when ionized in the mass spectrometer. Such an investigation has been performed, e.g., in the case of glycine, detected by the ROSINA mass spectrometer at 67P/C-G [81], which confirmed the earlier detection of glycine in the refractory dust returned from comet Wild 2 [23]. However, a unique identification may not always be possible in contrast to, e.g., spectroscopic observations of unambiguous rotational lines of the different isomers.

Table 2: Organic molecules identified in comets via remote and in situ methods. Column 1 lists the molecule in Hill notation, column 2 the name of the detected molecule. Indicated in *italic* font are species that may also be present in the form of a different isomer in the case of an ambiguous identification by in situ mass spectrometry. Column 3 lists the corresponding comet(s) or points to a reference, if a molecule has been detected at more than two comets, * tentative detection.

| **Sum formula** | **Name (*italic: additional isomers are possible*)** | **Comet(s) and references** |
|---|---|---|
| **C-bearing** | | |
| $CH_4$ | Methane | Several comets, e.g. Bockelée-Morvan, et al. [30] |
| $C_2H_2$ | Acetylene | Several comets, e.g. Bockelée-Morvan, et al. [30] |
| $C_2H_4$ | Ethylene | 1P/Halley, cf. Eberhardt [82] |
| $C_2H_6$ | Ethane | Several comets, e.g. Bockelée-Morvan, et al. [30] |
| $C_3H_8$ | Propane | 67P/C-G [72] |
| $C_4H_{10}$ | *Butane* | 67P/C-G [72] |
| $C_5H_{12}$ | *Pentane* | 67P/C-G [72] |
| $C_6H_6$ | Benzene | 67P/C-G [72] |
| $C_6H_{14}$ | *Hexane* | 67P/C-G [72] |
| $C_7H_8$ | Toluene | 67P/C-G [72] |
| $C_7H_{16}$ | *Heptane* | 67P/C-G [72] |
| $C_8H_{10}$ | *Xylene* | *67P/C-G [36] |
| $C_8H_{18}$ | *Octane* | *67P/C-G [36] |
| $C_{10}H_8$ | *Naphthalene* | *67P/C-G [36] |
| **CO-bearing** | | |
| $CH_2O$ | Formaldehyde | Several comets, e.g. Bockelée-Morvan, et al. [30] |
| $CH_4O$ | Methanol | Several comets, e.g. Bockelée-Morvan, et al. [30] |
| $C_2H_4O$ | *Acetaldehyde* | Several comets, e.g. Le Roy, et al. [32] |
| $C_2H_6O$ | *Ethanol* | Lovejoy [31], 67P/C-G [83] |
| $C_3H_6O$ | *Acetone*<br>*Propanal* | 67P/C-G [36, 84] |
| $C_3H_8O$ | *Propanol* | 67P/C-G [83] |
| $C_4H_{10}O$ | *Butanol* | 67P/C-G [83] |
| $C_5H_{12}O$ | *Pentanol* | 67P/C-G [83] |
| $CH_2O_2$ | *Formic acid* | Several comets, e.g. Le Roy, et al. [32] |
| $C_2H_4O_2$ | *Acetic acid*<br>*Methyl formate*<br>Glycoladehyde | Several comets, e.g. Le Roy, et al. [32]<br><br>Lovejoy [31] |
| $C_2H_6O_2$ | *Ethylene glycol* | Several comets, e.g. Le Roy, et al. [32] |
| $C_3H_6O_2$ | *Methyl acetate* | 67P/C-G [83] |
| $C_4H_8O$ | *Butanal* | 67P/C-G [83] |
| $C_7H_6O_2$ | *Benzoic acid* | *67P/C-G [36] |
| **CN-bearing** | | |
| CHN | Hydrogen cyanide<br>Hydrogen isocyanide | Several comets, e.g. Le Roy, et al. [32] |
| $CH_5N$ | Methylamine | 67P/C-G [81, 84] |
| $CH_3CN$ | *Acetonitrile* | Several comets, e.g. Le Roy, et al. [32] |
| $C_2N_2$ | Cyanogen | 67P/C-G [36] |
| $C_2H_7N$ | *Butylamine* | 67P/C-G [36] |
| $C_3HN$ | *Cyanoacetylene* | 67P/C-G [32]; Hale-Bopp [27] |
| $C_3H_9N$ | *Propylamine* | 67P/C-G [36] |
| **CNO-bearing** | | |
| CHNO | *Isocyanic acid* | Several comets, e.g. Le Roy, et al. [32] |
| $CH_3NO$ | *Methanamide* | Several comets, e.g. Le Roy, et al. [32] |
| $C_2H_5NO$ | *Acetamide* | 67P/C-G [84] |
| $C_2H_5NO_2$ | Glycine | Wild 2 [23], 67P/C-G [81] |
| **CS-bearing** | | |
| $CH_2S$ | Thioformaldehyde | 67P/C-G [67] |
| $CH_4S$ | Methanethiol | 67P/C-G [67] |
| $C_2H_6S$ | *Ethanethiol* | 67P/C-G [67] |
| $CH_4S_2$ | *Methyl hydrogen disulfide* | 67P/C-G [36] |
| **COS-bearing** | | |
| COS | Carbonyl sulfide | Several comets, e.g. Le Roy, et al. [32] |
| $CH_4OS$ | *Mercaptomethanol* | 67P/C-G [36] |
| $C_2H_6OS$ | *Mercaptoethanol* | 67P/C-G [36] |
| $C_3H_6OS$ | *Thiethane* | 67P/C-G [36] |

The dominant group of organics in terms of relative abundance are the C-H-bearing hydrocarbons [72], followed by C-H-O-compounds [83], C-H-N-, C-H-S-, and C-H-N-O-bearing molecules, and then even more complex compounds containing (combinations of) multiple O, N, and S atoms aside C and H. Furthermore, methyl isocyanate, $CH_3NCO$, has been suggested by Goesmann, et al. [84] but could not be confirmed by Altwegg, et al. [85] due to the absence of the corresponding fragment on mass/charge 56 Da/e in the mass spectrum of the Rosetta/Philae lander instrument COSAC. Measurements from the Philae lander mass spectrometer Ptolemy [86], just above the surface of comet 67P/C-G, suggested low amounts of aromatic compounds together with a lack of sulfur-bearing species and very low abundances of nitrogen-bearing molecules. Additionally, Wright, et al. [86], argued for the presence of polyoxymethylene (POM, i.e., polymerized formaldehyde) based on an alternating pattern of 14 and 16 Da/e corresponding to $-CH_2$ and -O units in the mass spectra of Ptolemy, respectively. POM, as the first polymerized molecule in space, was already proposed earlier from data of the PICCA (Positive Ion Cluster Composition Analyzer) instrument on board of ESA's Giotto mission passing comet 1P/Halley [87]. POM also forms easily under astrophysical conditions [88] and can be sputtered off a comet-like surface [89] where it may serve as a distributed source of formaldehyde [90], $H_2CO$, and carbon monoxide, CO. However, mass spectra of gaseous POM look different when the compound decomposes upon thermal desorption. This leads to a strong formaldehyde signal whereas the dimer, $(CH_2O)_2$, is already suppressed by 3 orders of magnitude. Altwegg, et al. [85] thus suggested that the corresponding 14 and 16 Da/e repetitive pattern in the Ptolemy mass spectra resulted from the occurrence of abundant C-H-O-bearing species (cf. Schuhmann, et al. [83]) and the associated $CH_2$- and O-fragments. Furthermore, the pronounced signal on mass/charge 91 Da/e corresponded to the $C_7H_7$ fragment of toluene, $C_7H_8$, detected on a regular basis in the coma of 67P/C-G by the Rosetta ROSINA instrument.

Table 3: Organic compounds detected in comets sorted by the number of C atoms and then by the molecular mass from Table 2, *italic: COMs*, * tentative detection.

| **Number of C-atoms** | | | | | | | | | |
|---|---|---|---|---|---|---|---|---|---|
| **$C_1$** | **$C_2$** | **$C_3$** | **$C_4$** | **$C_5$** | **$C_6$** | **$C_7$** | **$C_8$** | **$C_9$** | **$C_{10}$** |
| $CH_4$ | $C_2H_2$ | *$C_3H_8$* | *$C_4H_{10}$* | *$C_5H_{12}$* | *$C_6H_6$* | *$C_7H_8$* | *$C_8H_{10}$ | | *$C_{10}H_8$ |
| CHN | *$C_2H_4$* | $C_3HN$ | *$C_4H_8O$* | *$C_5H_{12}O$* | *$C_6H_{14}$* | *$C_7H_{16}$* | *$C_8H_{18}$ | | |
| $CH_2O$ | *$C_2H_6$* | *$C_3H_6O$* | *$C_4H_{10}O$* | | | *$C_7H_6O_2$ | | | |
| *$CH_5N$* | *$C_2H_3N$* | *$C_3H_9N$* | | | | | | | |
| *$CH_4O$* | *$C_2H_4O$* | *$C_3H_8O$* | | | | | | | |
| CHNO | *$C_2H_7N$* | *$C_3H_6O_2$* | | | | | | | |
| *$CH_3NO$* | *$C_2H_6O$* | *$C_3H_6OS$* | | | | | | | |
| $CH_2S$ | $C_2N_2$ | | | | | | | | |
| $CH_2O_2$ | *$C_2H_5NO$* | | | | | | | | |
| *$CH_4S$* | *$C_2H_4O_2$* | | | | | | | | |
| $CH_3Cl$ | *$C_2H_6S$* | | | | | | | | |
| COS | *$C_2H_6O_2$* | | | | | | | | |
| *$CH_4OS$* | *$C_2H_5NO_2$* | | | | | | | | |
| *$CH_4S_2$* | *$C_2H_6OS$* | | | | | | | | |

Table 3 lists the molecules from Table 2, sorted by the number of carbon atoms. COMs, which make up 37 of the 72 detected cometary species from Table 1, are marked in *italic* font. The list contains species of high volatility [91] such as methane, $CH_4$, as well as less volatile molecules that do not efficiently sublimate at comet surface temperatures < 200 K [63], but rather from the dust grains in the coma, which, depending on the size, can be up to several hundred degrees warmer at comparable heliocentric distances [64]. An example of a less volatile molecule is glycine, for which the density followed a more shallow dependence on the cometocentric distance compared the typical $r^{-2}$-dependence expected for a molecule being released from the nucleus [81]. For more information on the definition and nature of distributed sources of gas in the coma we refer to Cottin and Fray [92]. Not listed here is the large number of unsaturated hydrocarbons, which are present in the coma [72] , as it is very difficult to identify which ones are parent molecules and which ones are not. Furthermore, some of them may well originate from the more refractory organic components of 67P/C-G, as shown by Altwegg, et al. [85] in the case of dust impact observations later in the mission.

## 2.3. Organics in the refractory phase

Aside from the volatile organic molecules discussed in this work, we shall briefly touch on the abundant amounts of organics that have been observed in the refractory component, e.g., in cometary dust of various comets including comet 1P/Halley by the PUMA instrument [93] on the Soviet Vega 1 mission [49] and in the returned dust grains from comet Wild 2 by NASA's Stardust spacecraft [22, 28]. Organics have also been observed remotely on the surface of comets: measurements by VIRTIS aboard ESA's Rosetta spacecraft [94] suggested that a top layer of a dark refractory polyaromatic and aliphatic carbonaceous material was responsible for the observed low albedo [95-96] of 67P/C-G. The surfaces of comets are irradiated by solar photons, cosmic rays, and, at times of low activity, even solar wind (energetic) particles make it to the nucleus [97-99]. These may form such surface compounds. Erosion processes, on the other hand, refresh the surface layer releasing some of these materials into the coma.

The lack of signatures of typical hydrated minerals, such as phyllosilicates, is evident in the mass spectra obtained with the COSIMA dust mass spectrometer [69] and suggests a dry history for comet 67P/C-G. Investigations based on COSIMA data furthermore reveal that approximately 55% of the dust mass is made up of a mineral phase and the remaining 45% of organics [69] resembling Insoluble Organic Matter (IOM) found in carbonaceous chondrites and bound in large macromolecular compounds [100]. Bardyn, et al. [69] reported some differences compared to the IOM found in carbonaceous chondrites, in particular, the elevated H/C ratio potentially highlighting the more primitive origin of cometary organics with respect to chondrites.

Composition measurements of the returned dust grains from comet Wild 2 by NASA's Stardust mission revealed a whole plethora of organics: some similar to interplanetary dust particles and carbonaceous chondrites, yet others depleted in aromatics [22]. On average, they are rich in oxygen and nitrogen compared to organics in meteorites [22]. A later re-analysis of the Stardust grains also revealed the presence of glycine [23], the simplest amino acid. Stardust samples also indicate abundant amounts of relatively labile organics may have been lost into the aerogel upon impact [22], possibly linking the measurements of the organics sublimating from grains in the coma [85] and the insoluble macromolecular phase reported by Fray, et al. [100] for 67P/C-G.

# 3. Origin of organic molecules in comets

The origin of cometary material is still a subject of debate. The detection and abundance of several molecules and their isotopologues suggest that at least part of the cometary ices in 67P/C-G is of prestellar origin [77]; and that subsequent processing has been marginal [36] or limited to the amorphous-to-crystalline ice phase transition of the ices with subsequent re-trapping of the volatiles [101]. As a consequence, this would also apply to the (complex) organic molecules in the comet. The following section discusses some of the relevant isotopic measurements, followed by a discussion based on the volatile inventory present in comet 67P/C-G.

## 3.1. Formation in the dark cold prestellar core – isotopic record

Some of the refractory material returned by NASA's Stardust mission [28] is the result of high temperature processing [102], either close to the Sun or possibly by shocks. On the other hand, the same samples also contain organics with deuterium and nitrogen-15 excesses [22]. Elevated D/H ratios in various molecules are commonly observed star-forming regions and are interpreted to be a consequence of grain-surface chemistry at very low (~10 K) dust temperatures during the prestellar core phase [103-106]. Hence, an observed excess of D in cometary material may indicate a prestellar or protostellar heritage. Lower D/H ratios, on the other hand, may be found the inner warm regions of the Solar Nebula that were heated by the Sun during the formation of our Solar System. Consequently, cometary refractories are likely to be a mixture of materials from a wide range of heliocentric distances that have been subject to radial transport before incorporation into the comet.

Excesses of deuterium [44] and nitrogen-15 [107] are also commonly observed in the volatile molecules of comets. In particular, the very similar $^{14}N/^{15}N$ ratios, of 114 ± 3 on average, measured in cometary HCN, CN, and $NH_2$ differ consistently from the $^{14}N/^{15}N = 441 \pm 6$ ratio of the Sun [108] (i.e., the bulk of our solar system) and also from the terrestrial [109] ratio of $^{14}N/^{15}N = 273 \pm 1$. Furthermore, nitrile and amine functional groups in prestellar cores exhibit different $^{14}N/^{15}N$ ratios [110], which has been interpreted as direct evidence for multiple reservoirs of nitrogen [107], one deriving from atomic and the other from molecular nitrogen [111]. Numerical models, however, seem to have difficulty reproducing, for instance, the strong $^{15}N$ depletions observed in the prestellar core L1544 [112]. On the basis of the proposed reaction scheme by Hily-Blant, et al. [111], the two separate reservoirs are also challenged by the measurements at comets: HCN and $NH_2$ (from $NH_3$), formed from the two distinct atomic and molecular reservoirs, respectively, show very consistent $^{14}N/^{15}N$ ratios in comets [107]. Key measurements of $^{14}N/^{15}N$ in $NH_3$, NO (intermediate species relating the two reservoirs [111]), and $N_2$ in a comet are therefore required.

The spread among comets in the water D/H ratio, on the other hand, is larger and ranges from the terrestrial-like ratio of $(1.61 \pm 0.24) \cdot 10^{-4}$ in Jupiter-family comet Hartley 2 [43] to the 4-times enriched value in the Oort-cloud comet C/2012 F6 (Lemmon) [113] relative to the terrestrial Vienna Standard Mean Ocean Water (VSMOW) D/H ratio of $(1.558 \pm 0.001) \cdot 10^{-4}$. In particular, no apparent distinction between Oort cloud comets and Jupiter-family comets has been found. Earlier, it was hypothesized that both populations could have formed at distinct distances from the Sun; therefore, inheriting different D/H ratios. However, the finding of a more than 3-times elevated D/H ratio in comet 67P/C-G compared to VSMOW suggests that comets form over a wide range of heliocentric distances before being expelled to the Kuiper-belt, the Scattered Disk, or the Oort Cloud, where they are found today [44]. Furthermore, the recent measurement of a terrestrial-like D/H ratio in the water of comet 46P/Wirtanen [114], suggests a correlation to a comet's activity in relation to its nucleus size as opposed to its dynamical origin.

An elevated $[HDO]/[H_2O]$ ratio ($=2 \cdot D/H$) alone may not be sufficient evidence for inheritance of water ice from the prestellar core. Indeed, the cold envelope of NGC1333 IRAS2A has a 70 times higher D/H ratio obtained from $HDO/H_2O$ compared to that of comet 67P/C-G [115], which would require very little of this highly deuterated water to mixed with weakly deuterated water (which could potentially be residing in the warmer parts of the Solar Nebula) to obtain the water D/H ratio of 67P/C-G [44]. As an alternative, Furuya, et al. [116] proposed to compare the D/H ratios of singly and doubly deuterated water. This was motivated by Coutens, et al. [117], who reported an approximately 7-times higher $[D_2O]/[HDO]$ (=½ · D/H) abundance ratio compared to $HDO/H_2O$ in the gas phase around the low mass protostar NGC 1333 IRAS2A. Statistically, one expects: $[D_2O]/[HDO] / [HDO]/[H_2O]$ = (½ · D/H) / (2 · D/H) = ¼. Two possible mechanisms had been envisioned to explain this enrichment in doubly deuterated water. The first is mixing of D-enriched water from the prestellar core with freshly formed, D-depleted water in the warmer regions ($\gtrsim$ 230 K) close to the protostar [117]. The second is a 2-stage model, where water is formed at moderate temperatures on icy grains and then covered by a HDO- and $D_2O$-enriched layer, as the core cools and deuteration reactions become more efficient [116]. Both theories require that at least parts, if not the majority, of the ices have remained unaltered since their formation in the prestellar core. The D/H enrichment in water is sensitive to temperature: Furuya, et al. [116] started from a 10 K pre collapse phase temperature resulting in a $HDO/H_2O$ ratio on the order of $10^{-3}$, well above typical values found at comets [114]. Aside from the addition of D-depleted water from warmer regions discussed above, this may alternatively point at a slightly higher temperature of the core feeding our Solar System's comets, as suggested by a model [118] reproducing the $O_2/H_2O$ ratio found in 67P/C-G [70], which will be discussed in the next section.

At comet 67P/C-G, an even higher D/H ratio in the doubly over the singly deuterated water [73] was obtained, i.e., $[D_2O/HDO] / [HDO/H_2O]$ = 17 instead of ¼. Hence, we expect the amounts of ices inherited from the prestellar core to be sizeable. Here, it is worthwhile to account for the D-H isotope exchange reactions in amorphous water ices, which are representative of water ice that has never undergone sublimation. Laboratory experiments [119-121] show that hydrogen-deuterium isotope exchange reactions occur on time-scales of 10,000 yrs at temperatures $\gtrsim$ 70 K. This obviously must have been prevented in 67P/C-G based on the measured ratios. On the other hand, inhibition of H-D exchange reactions may also be a consequence of the 2-stage water formation model by Furuya, et al. [116] given that the $H_2O$ on the grain is separate from the layer containing HDO and $D_2O$ and hence, isotope exchange is limited to the boundary between these layers. However, the conclusion that at least parts of the ice remained in its pristine prestellar form still stands.

Furthermore, measurements at comet 67P/C-G indicate deviations relative to the respective standards in the isotopic ratios of (i) oxygen [74] in $H_2O$ ($\delta^{17}O = (+206 \pm 94)$‰ and $\delta^{18}O = (+122 \pm 90)$‰ deviation from VMSOW), (ii) carbon [122] in $CO_2$ ($\delta^{13}C = (+59 \pm 48)$‰ deviation from terrestrial Pee Dee Belemnite), (iii) sulfur [123] with a mean deviation of $\delta^{33}S = (-302 \pm 29)$‰ and $\delta^{34}S = (-41 \pm 17)$‰ from Vienna-Canyon Diablo Troilite in $H_2S$, OCS, and $CS_2$, (iv) silicon sputtered off the comet's surface by the solar-wind [75] ($\delta^{29}Si = (-148 \pm 98)$‰ and $\delta^{30}Si = (-214 \pm 115)$‰ deviation from NIST NBS 28 quarz sand), and (v) noble gases, with xenon [71] (enriched in $^{129}Xe$ and depleted in both $^{134}Xe$ and $^{136}Xe$ compared to the solar wind) and possibly krypton [65] ($\delta^{83}Kr = (-81 \pm 34)$‰ compared to the solar wind). On the other hand, $^{12}C/^{13}C$ in CO, $^{16}O/^{18}O$ in $CO_2$, $^{35}Cl/^{37}Cl$, $^{79}Br/^{81}Br$, ($^{80}Kr/^{84}Kr$), $^{82}Kr/^{84}Kr$, $^{86}Kr/^{84}Kr$, $^{128}Xe/^{132}Xe$, $^{130}Xe/^{132}Xe$, and $^{131}Xe/^{132}Xe$ are all indistinguishable from their corresponding standard at the 1-$\sigma$ level. Given these isotopic variations of a given element among different molecules and noticeable deviations from the solar isotopic composition, these results point to a non-homogeneously mixed protosolar nebula. Also, a comparison of the isotope ratios in 67P/C-G and in the primitive solar system materials indicates that the comet is especially primitive and contains, at least in parts, un- or little processed material of presolar origin [77].

## 3.2. Formation in the dark cold prestellar core – molecular record

Abundant amounts [124] of highly volatile species [30] such as, e.g., $CH_4$ and CO confirm that 67P/C-G has never experienced temperatures above a few tens of K since its assembly/formation [125]. The absence of hydrated minerals [69], contrary to CI chondrites, is evidence that the comet has never been host to liquid water. Therefore, the material in 67P/C-G may never have been part of a larger object, in which processing in the presence of liquid water could have occurred prior to the cometary body emerging upon an impact on the larger parent object.

Several of the molecules known to be present in comets have also been observed in the ISM. Table 1 lists the molecules for which at least one isomer has been observed in the ISM [62], showing a significant overlap in the detected species, i.e., 41 out of the 72 cometary molecules from Table 1 are also known to be present in the ISM. Furthermore, and as underlined above, there may be many more molecules in common, which have merely not yet been identified due to a lack of spectroscopic data necessary for their identification with powerful facilities such as ALMA.

Bockelée-Morvan, et al. [27] compared the relative volatile abundances of a suite of CHO-compounds, as well as nitrogen- and sulfur-bearing species, detected in comet Hale-Bopp with the corresponding ratios of several star-forming regions. In particular, CHO- and nitrogen-bearing species showed evidence for a link between comets and the ISM, as cross-correlations between the two populations appeared to be in agreement. For instance, Bockelée-Morvan, et al. [27] reported [$CH_3OH$]/[HCN] ratios towards the high-mass protostars G34.3+0.15 and W3($H_2O$), as well as the low-mass embedded protostar L1157 to be within a factor of 3, and in the Orion Hot Core and the Compact Ridge to be within a factor of 5 - 10, from those of comet Hale-Bopp. Figure 1 represents a similar comparison between comet 67P/C-G and the same star-forming regions with O-bearing compounds on the left, N-bearing species in the middle, and S-bearing molecules on the right. The match between sulfur-bearing species is less pronounced for both, Hale-Bopp [27] and 67P/C-G (right panel Figure 1), most likely due to more pronounced gas-phase chemistry occurring in the hotter, more dense envelopes of high-mass protostars that are well-shielded from the internal protostellar UV field, the so-called hot cores, in comparison to the physical conditions in the environs of the early Sun. In particular, the [$CH_2S$]/[COS] ratio shows 3 orders of magnitude variation among the probed star-forming regions. Hence, some of the protostellar targets in this comparison may be less relevant for the expected physical conditions of our early solar system. Drozdovskaya, et al. [42] carried out an in-depth investigation comparing the low-mass protostar

IRAS 16293-2422B with comet 67P/C-G. They showed that a comparison to a source more comparable to our proto-Sun does improve the match for the sulfur-bearing molecules.

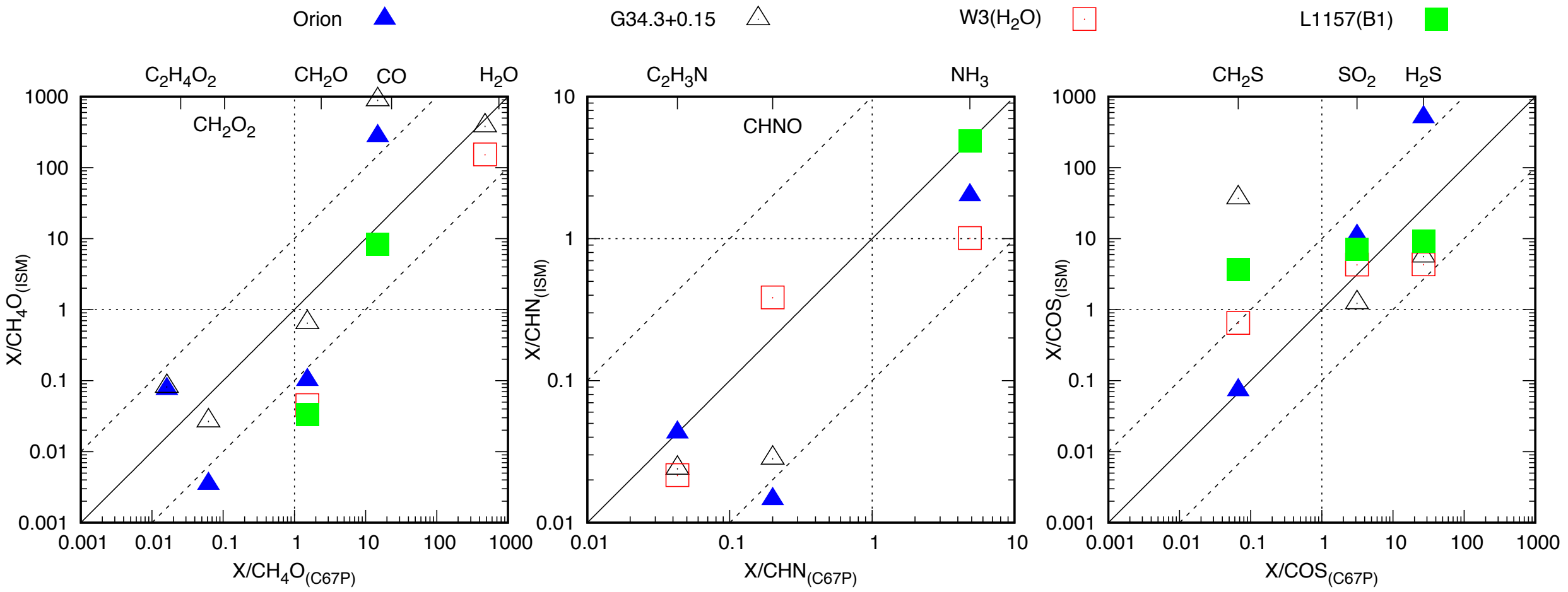

Figure 1: Relative abundances of oxygenated compounds w.r.t. $CH_4O$ (left), nitrogen-bearing species w.r.t. CHN (middle), and sulfur-bearing molecules w.r.t. COS (right) in several star-forming regions [27] versus those in the bulk of comet 67P/C-G [124]. The diagonal solid black lines denote equal mixing, the diagonal dashed lines enclose differences up to a factor of ten, and vertical and horizontal lines mark the reference compounds $CH_4O$, CHN, and COS. For the case of 67P/C-G, some of the indicated molecules may be present in the form of different isomers.

Quite unexpected were the high amounts of molecular oxygen measured in situ [70] in the coma of comet 67P/C-G and inferred from remote sensing observations [126]. Also, the later re-analysis of Neutral Mass Spectrometer [52] data from ESA's Giotto mission [50] suggested $O_2$ to be present in comet 1P/Halley's coma [127]. The measurements also showed a good correlation of $O_2$ with $H_2O$, despite the very different sublimation temperatures of the corresponding pure ices [91]. This indicates that the $O_2$ is trapped within the bulk $H_2O$ ice matrix of comet 67P/C-G.
$O_2$ is known to form in the irradiated environments of the water-ice covered moons of Jupiter [128] and in Saturn's icy rings [129]. $O_2$ (although at lower abundances [70], possibly due to an observational bias related to the desorption temperature of $O_2$ from amorphous ice [130]), $HO_2$, and $H_2O_2$ have also been observed as gases near low-mass protostars [131-134]. Consequently, energetic processing of prestellar ices by forming protostars before their incorporation into a comet has been proposed as a possible $O_2$ formation mechanism [70, 135]. This is in agreement with the rather good correlation of $O_2$ with $H_2O$ observed at 67P/C-G, but may be at odds with the low abundance of other typical radiolysis products $HO_2$ and $H_2O_2$ (and the absence of $O_3$) at 67P/C-G [70]. Mousis, et al. [135] argued that the formation of $O_2$ occurred before the accretion of the comet [70] and irradiation and of icy grains is not fast enough in the protosolar nebula but has to occur earlier in low-density environments such as molecular clouds to be efficient [135].

An alternative primordial formation scenario is based on oxygen atom recombinations on the surface of water ice-covered dust grains under slightly denser ($n_H \gtrsim 10^5$ $cm^{-3}$) and warmer (~20 K) conditions of prestellar cores [118]. In this model, species such as $N_2$ and CO form or are trapped later and hence exhibit a lower correlation with $O_2$ and $H_2O$, similar to what is observed at 67P/C-G [136].

Aside from a primordial origin of $O_2$, two in situ formation scenarios have been proposed (cf. review by Luspay-Kuti, et al. [137]). The first is based on the dismutation of hydrogen peroxide, $H_2O_2$, during the sublimation process [138]. However, the low abundance of $H_2O_2$ measured at 67P/C-G [70] requires dismutation to be very efficient. The second formation mechanism invokes Eley-Rideal reactions of energetic water ions with an oxidized surface [139]. This explanation suffers from the poor correlation of energetic water ions with both neutral $H_2O$ and $O_2$ and, furthermore, too low water ion fluxes [140].

Formation models of complex organics in prestellar cores, protostellar envelopes, and protoplanetary disks predict large amounts of unsaturated species [141]. The organics in comets also contain sizeable amounts of unsaturated CH- and CHO- bearing molecules (cf. section 2.2), as shown by ROSINA measurements of coma gases at 67P/C-G: Schuhmann, et al. [72] showed that for many CH-bearing molecules most of the signal remained unaccounted for, even after subtraction of fragments produced in the ion source of saturated CH- bearing compounds and cycloalkanes. A similar picture arises for CHO- bearing compounds [83].

In the refractory organic phase, carbonaceous matter, often attributed to a prestellar and protostellar origin [142], and highly unsaturated compounds have been observed at comet 1P/Halley by the PUMA instrument on the Vega 1 mission [93]. The Rosetta/COSIMA instrument detected predominantly $C^+$, $CH^+$, $CH_2^+$, $CH_3^+$, and $C_2H_3^+$ fragment ions released from large macromolecular matter [100] by the instrument's pulsed 8 keV indium beam.

## 3.3. Formation of cometary volatiles in prestellar cores and star-forming regions

Isotopic ratios and the chemical inventory of cometary volatiles appear to be defined at least to some degree during the preceding evolutionary phases of our Solar System, likely going back as far as the protostellar and prestellar stages, as supported by the evidence presented in Sections 3.1 and 3.2 This implies that cometary ices are assembled from atoms along the same chemical pathways as the ices studied by astrochemistry under interstellar physical conditions. The dominant cometary volatile molecule, water, is thought to be formed via gas-phase reactions stemming from an ion-molecule reaction of $OH^+$ with $H_2$ at low temperatures and through hydrogenation of O, $O_2$, and $O_3$ on the surfaces of dust grains in cold, dense prestellar cores [143]. Other high-temperature routes exist as well; however, these are likely to be less relevant for the cometary ices, which are thought to remain cold during the entire duration of the assembly of the cometary body. The formation of $CO_2$ from CO+OH competes with the formation of $H_2O$ from OH+H on grain surfaces [144]. Under dark, non-irradiated conditions, the barrierless hydrogenation reaction resulting in $H_2O$ dominates. However, upon exposure to UV radiation, physicochemical models have shown that $CO_2$ may start to efficiently form via CO+OH at the expense of water due to its higher photostability [9]. CO is formed in the gas phase already during the translucent cloud phase of the ISM, alongside $H_2O$. It forms during the transition of $C^+$ to C and to CO as one considers deeper zones of a photodissociation region (e.g., Glover, et al. [145]). $N_2$ and $NH_3$ are presumably the main carriers of nitrogen in the earliest phases of molecular clouds, stemming from neutral-neutral and ion-molecule reactions in the gas phase, respectively [146]. The abundances of $NH_3$ and $CH_4$ are further boosted via grain-surface reactions once hydrogenation becomes efficient [147]. The chemistry of S-bearing molecules, including $H_2S$, remains debated and less constrained (e.g., Vidal, et al. [148]).

The routes towards more complex O-, N- and S-bearing molecules are much more intertwined. Methanol is considered to be the key parents species for further chemical complexity as it can serve as abundant source of the $CH_3$ and OH functional groups. In the gas, it forms through a two-step process: radiative association [149], followed by dissociative recombination, which has been shown experimentally to be inefficient [150] and further suppressed by other competing ion-molecule reactions with larger rate coefficients [151]. Grain-surface chemistry does provide an efficient pathway towards $CH_3OH$ via sequential hydrogenation of CO, as suggested by theory [152] and verified with experiments [153]. If UV is available, either from the forming protostar or the interaction of cosmic rays with $H_2$, then methanol can also form via radical-radical association reactions on the surface ($CH_3$+OH). Additional formation pathways on grain surfaces through, e.g., radiolysis by ~1 MeV protons, have also been suggested in the literature [154-155]. Observations indicate that $CH_3OH$ concentrations can vary by an order of magnitude between star-forming regions, which may be a measure of the UV flux and/or dust temperatures around protostars [156]. Higher chemical complexity forms via a combination of hydrogenation and radical-radical reactions depending on the physical conditions (e.g., Hasegawa, et al. [141,] Garrod [157,] Ioppolo, et al. [158,] Fedoseev, et al. [159]).

Also the chemistry of S-bearing molecules remains debated and less constrained. Models of prestellar cores are suggesting the distribution of elemental sulfur across a large set of molecules including $H_2S$, HS, CS, $H_2CS$, OCS, and more (Vidal, et al. [148]; Laas and Caselli [160]). This may explain why $H_2S$ ice, initially thought to be the primary S-carrier, has not been detected directly with infrared observations at an upper limit of ~1% relative to water (Smith [161]; Boogert, et al. [1]). The gas-phase abundance of $H_2S$ towards the low-mass protostar IRAS16293-2422B has been shown to differ significantly from its high abundance in comet 67P/C-G [42], which may suggest it being heavily depending on the collapse physics of each and every star-forming region [162].

# 4. Biological relevance

The assembly of prebiotic molecules may have already started in space [163]. Comets contain sizeable amounts of volatile species with the elements C, H, N, O, P, and S, which make up the vast majority of the biomolecules on Earth. Through impacts, some of them potentially find their way to the inner planets, including Earth, where they can contribute refractories to the crust [163], water to the oceans [43], volatiles to the atmosphere [12], and a suite of organic molecules relevant for pre-biotic chemistry [164] to the planet. The major portion of this delivery was suggested to have occurred during the Late Heavy Bombardment (LHB) from $4.5 \cdot 10^9$ to $3.8 \cdot 10^9$ years ago, just around the time of the emergence of life on Earth [164].

Table 4: Mass fraction relative to $H_2O$ in the bulk ice of comet 67P/C-G based on Rubin, et al. [124] with references therein [65, 67, 72, 83, 165]. *Italic*: species considered to derive the organic mass in the cometary ices. Listed are 1-σ error bars except for HF, HCl, HBr, and PO, for which the observed relative ranges are given.

| **Molecule** | **Mass fraction w.r.t. $H_2O$** | **Molecule** | **Mass fraction w.r.t. $H_2O$** |
|---|---|---|---|
| $H_2O$ | 1 | *HNCO* | 0.00065 ± 0.00038 |
| $CO_2$ | 0.12 ± 0.03 | *$CH_3NO$* | 0.00010 ± 0.00006 |
| CO | 0.048 ± 0.015 | *$CH_3CN$* | 0.00013 ± 0.00008 |
| $O_2$ | 0.054 ± 0.020 | *$HC_3N$* | 0.000011 ± 0.000007 |
| *$CH_4$* | 0.0031 ± 0.0006 | $H_2S$ | 0.021 ± 0.009 |
| *$C_2H_6$* | 0.0049 ± 0.0010 | *OCS* | $0.0014^{+0.0027}_{-0.0007}$ |
| *$C_3H_8$* | 0.00044 ± 0.00009 | SO | $0.0019^{+0.0038}_{-0.0010}$ |
| *$C_6H_6$* | 0.000030 ± 0.000006 | $SO_2$ | $0.0045^{+0.0090}_{-0.0023}$ |
| *$C_7H_8$* | 0.00031 ± 0.00006 | $CS_2$ | $0.00024^{+0.00048}_{-0.00012}$ |
| *$CH_3OH$* | 0.0037 ± 0.0011 | $S_2$ | $0.000070^{+0.000140}_{-0.000037}$ |
| *$C_2H_5OH$* | 0.0010 ± 0.0006 | *$H_2CS$* | $0.000069^{+0.000148}_{-0.000062}$ |
| *$C_3H_6O$* | 0.00015 ± 0.00008 | S | 0.0081 ± 0.0065 |
| *$H_2CO$* | 0.0053 ± 0.0017 | *$CH_3SH$* | $0.0010^{+0.0021}_{-0.0007}$ |
| *$CH_3CHO$* | 0.0011 ± 0.0004 | *$CH_3CH_2SH$* *$CH_3SCH_3$* | $0.000020^{+0.000042}_{-0.000017}$ |
| *HCOOH* | 0.00033 ± 0.00020 | Ar | 0.000012 ± 0.000004 |
| *$(CH_2OH)_2$* | 0.00037 ± 0.00023 | Kr | 0.0000023 ± 0.0000010 |
| *$CH_3COOH$* | 0.00011 ± 0.00007 | Xe | 0.0000018 ± 0.0000008 |
| *$C_3H_6O_2$* | 0.000086 ± 0.000027 | HF | $0.00012^{+0.00041}_{-0.00009}$ |
| *$C_4H_8O$* | 0.00040 ± 0.00013 | HCl | $0.00029^{+0.00091}_{-0.00024}$ |
| $NH_3$ | 0.0063 ± 0.0019 | HBr | $0.000013^{+0.000024}_{-0.000008}$ |
| $N_2$ | 0.0014 ± 0.0004 | PO | $0.00029^{+0.00057}_{-0.00014}$ |
| *HCN* | 0.0021 ± 0.0006 | | |

Table 4 lists relative abundances by mass of a suite of major and minor volatile species with respect to cometary $H_2O$ in the bulk ice of comet 67P/C-G derived from Rubin, et al. [124]. The bulk ice is dominated by water, followed by carbon dioxide, molecular oxygen, and carbon monoxide. While the relative abundances differ among comets [30], we will base our computations on this list, which is the most comprehensive of any comet to date.

## 4.1. Transport to Earth

The role of comets as a major source of the terrestrial water have been discussed after an Earth-like D/H ratio has been detected in the Jupiter-family comet 103P/Hartley 2 by the Herschel Space Observatory[43]. All earlier D/H observations in cometary water were obtained for Oort cloud comets and exhibited deuterium enrichment incompatible with the Vienna Standard Mean Ocean Water (VSMOW) value of $D/H = (1.558 \pm 0.001) \cdot 10^{-4}$. However, recent measurements of elevated D/H ratios in the water of the Jupiter-family comet 67P/C-G [44] and the Oort-cloud comet C/2012 F6 (Lemmon) [113] have again put doubt on this conjecture.

Measurements of the isotopic composition of noble gases mixed into the water ice sublimating from comet 67P/C-G have yielded crucial insights into the origin of volatile elements on terrestrial planets [71]. The ROSINA mass spectrometry system on-board the Rosetta spacecraft permitted the Ar, Kr, and Xe abundances in cometary ice to be determined [65]. The detection of noble gases in comet 67P/C-G indicates the possibility of a cometary contribution to the terrestrial atmosphere. In particular, the deficits in heavy $^{134}Xe$ and $^{136}Xe$ isotopes measured within comet 67P/C-G relative to solar matched that of the Earth's primordial atmospheric component [71]. In order to replicate the ancient Earth's atmosphere, (22 ± 5)% of Xe in the terrestrial atmosphere is required to be of 67P/C-G-like origin. In comparison, on the basis of the relative abundances of water versus noble gases (cf. Table 4), such a contribution would deliver less than 1% of the water to the terrestrial oceans, resulting in only a minor water D/H enhancement [12].

A cometary contribution to the terrestrial atmosphere inventory had long been suspected. Interestingly, the Kr/Xe ratio of the terrestrial atmosphere ($^{84}Kr/^{132}Xe = 27$; Porcelli, et al. [166]) is close to the solar ratio ($^{84}Kr/^{132}Xe = 29 \pm 8$; Wieler [167]), but different from chondritic values ($^{84}Kr/^{132}Xe < 1.3$; Busemann, et al. [168]). The chondritic signature of the heavy noble gases in the Earth's mantle led Holland, et al. [169] to argue that mantle degassing could not generate the modern atmosphere, and that an additional high $^{84}Kr/^{132}Xe$ source was therefore required to buffer the atmospheric $^{84}Kr/^{132}Xe$ towards solar values. Laboratory experiments of noble gas trapping in amorphous water ice predict enrichment in Kr relative to Xe in cometary ice with respect to the starting composition [170]. This suggests that cometary ices condensing from a reservoir with a solar isotopic composition could have provided such a high $^{84}Kr/^{132}Xe$ source. Likewise, Ar/Xe-Kr/Xe correlations between terrestrial and Martian samples [171] combined with laboratory analyses of noble gases trapping in amorphous ice by Bar-Nun and Kleinfeld [172] permitted the formation temperature of comets contributing the volatile element budget of terrestrial planets in a late accretionary stage to be estimated (~50 K; Owen and Bar-Nun [173]). However, noble gas abundances measured by ROSINA ($^{84}Kr/^{132}Xe = 3.7 \pm 0.9$; Rubin, et al. [65]) are in fact similar to chondritic values. The high $^{84}Kr/^{132}Xe$ of the terrestrial atmosphere may therefore be accounted for by Xe-specific loss to space through time [174]. In particular, the competition between burial trapping and equilibrium trapping of noble gases in amorphous ice, mainly relying on the temperature and rate of ice deposition, should ultimately be accounted for when determining the source of trapped volatiles during planet formation [175].

Unlike in the atmosphere, primordial heavy noble gases detected in the terrestrial mantle are seemingly chondritic in origin[169, 176]. The absence of significant cometary contribution to the mantle noble gas inventory has been used to argue for a late arrival of comets to Earth, after the Moon-forming impact. Based on the age and size distribution of the lunar cratering record, it has been proposed that a cataclysmic spike in the material impacting the Moon, and presumably the Earth, occurred 700 Myr after the end of planet formation [177]. It has been proposed that this late spike in the material striking the inner planets is related to the rapid migration of the giant planets, which caused a destabilizing effect on materials outside of the orbit of the giant planets and resulted in the outer solar system material, potentially including comets, being injected into the inner solar system [13]. One promising avenue of investigation to further constrain the timing of the cometary supply to the Earth-Moon system is to determine whether or not cometary volatiles are preserved on the present-day Moon surface. Measurements of hydrogen [178] and xenon [179] isotopes indicate that the cometary volatiles may be present on the surface of the Moon. The composition measurements of the LCROSS (Lunar Crater Observation and Sensing Satellite) mission of a plume of volatiles, excavated by an impactor from an area in permanent shadow, revealed processed material of partly cometary and/or asteroidal origin [180]; however, more work is required to determine the contribution and timing of any cometary input to the Moon's surface. Whether cometary volatiles are a vital constituent of all the terrestrial planets remains to be seen. For example, the $C/N/^{36}Ar$ elemental abundance ratios and low $^{14}N/^{15}N$ measured in the Martian atmosphere are consistent with a late addition of cometary material to Mars [12]. However, the Xe isotopic signature of the Martian atmosphere does not require any input from cometary material and is best accounted for by a solar-derived origin of Martian heavy noble gases [181]. The role of cometary volatiles in forming the tentative Martian atmosphere therefore remains up for debate.

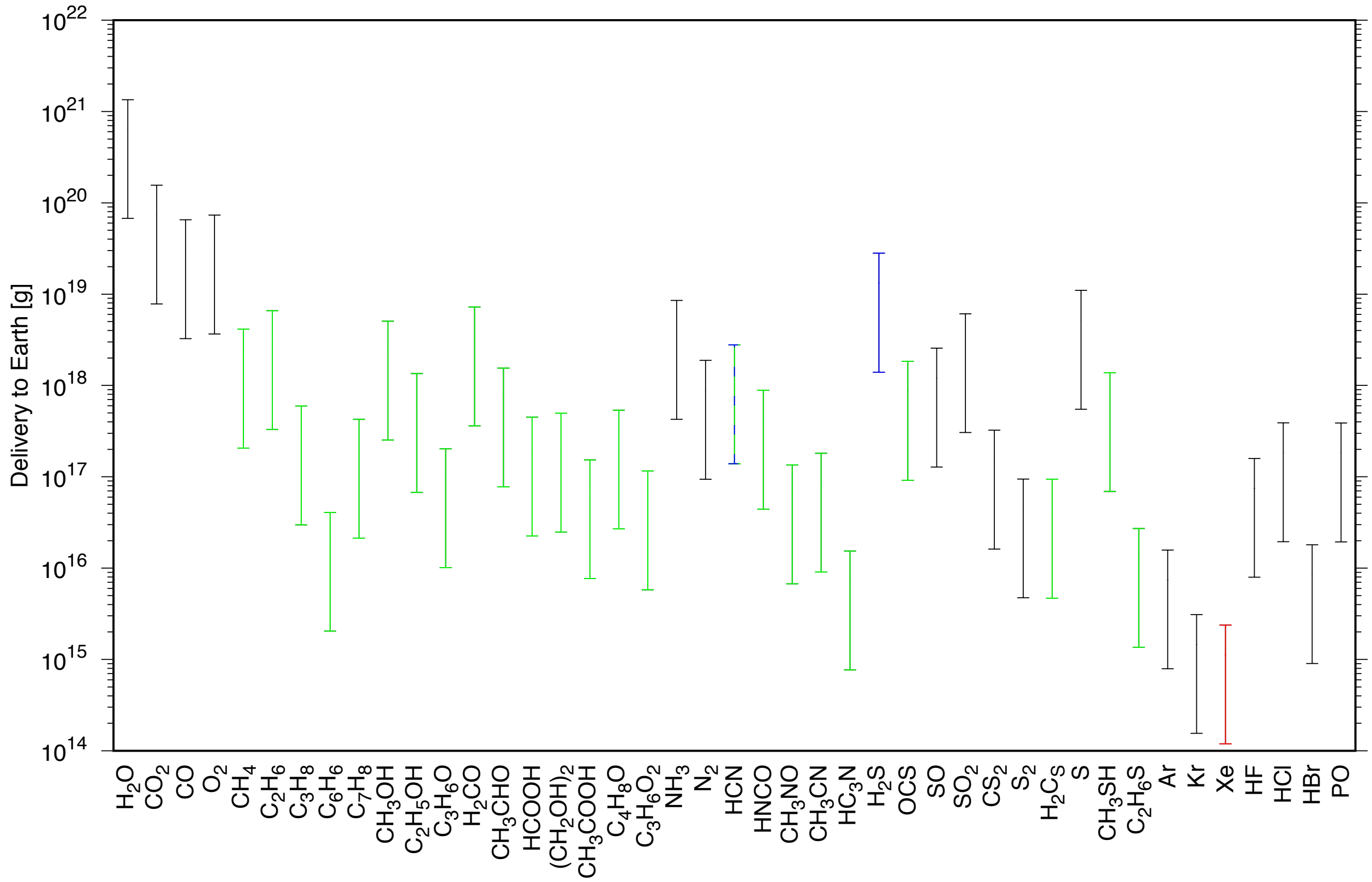

Figure 2: Mass delivery to Earth split among different molecules in [g]. Scaling is based on a 22% contribution of cometary xenon (red) to the terrestrial atmosphere and the relative mass fraction from Table 4. Organic species are indicated in green (cf. Table 4) and HCN as well $H_2S$ in blue (see text for more details).

Before and especially during the LHB [13], but also up to today [182-183], sizeable amounts of cometary and non-cometary materials have been delivered to Earth. Regarding comets, we can also attempt to calculate the amount of volatile species brought to Earth through impacts by the constraint set by terrestrial atmospheric Xe isotopes, which requires that comets contributed $(22 \pm 5)\%$ of the Xe residing in the Earth's atmosphere [71]. Based on Table 4, we computed the mass of individual volatile species delivered to a modern-day atmosphere [184] of $4.1 \cdot 10^{12}$ mol Xe and to an early atmosphere [185] containing 20-times more Xe (cf. Ozima and Podosek [186]), which corresponds to $1.2 \cdot 10^{14}$ to $2.4 \cdot 10^{15}$ g as indicated by the red bar in Figure 2. The resulting total mass of $H_2O$ corresponds to $6.8 \cdot 10^{19}$ to $1.4 \cdot 10^{21}$ g. This corresponds to a 0.045 to 0.9 ‰ contribution to the total ocean mass [12] of approximately $1.5 \cdot 10^{24}$ g and therefore, limits the enhancement in the ocean's D/H ratio to a few per mil due to the elevated D/H ratio in cometary water. In comparison, the NICE model [13] suggests a total of $\sim 1.8 \cdot 10^{23}$ g of cometary mass (refractories and volatiles) was delivered to Earth during the LHB. However, this number is considered to be an upper limit for the delivery of water to Earth based on cometary D/H ratios [187]. Furthermore, the total mass of organics can be estimated. For this purpose, we include the list of molecules from Figure 2 indicated in green (*italic* in Table 4). The resulting mass range of $1.7 \cdot 10^{18}$ to $3.4 \cdot 10^{19}$ g represents several hundred times the total oceanic biomass [164] of $3 \cdot 10^{15}$ g or is similar to the total terrestrial biomass [188] of $2 \cdot 10^{18}$ g. This does not however, necessarily mean that all biomolecules are of cometary origin, but that the contribution may have been sizeable.

The number of corresponding cometary impacts ranges between 17,000 – 350,000 when derived based upon 67P/C-G's mass [189] of approximately $10^{13}$ kg and assuming a dust-to-ice ratio of 1 inside the nucleus (cf. Läuter, et al. [190]). If referenced to $2.2 \cdot 10^{14}$ kg, the mass of comet 1P/Halley [191], the number of impacts has to be divided by 22 when assuming the same dust-to-ice ratio. The corresponding number of impacts for different dust-to-ice ratios and impactor masses can easily be derived from these numbers. Importantly, note that the total amount of organics in the bulk of comet 67P/C-G is entirely dominated by the refractory component, independent of any reasonable dust/ice ratio inside the bulk of the comet, and only a fraction of the material bound in the solids may be relevant for pre-biotic chemistry. This simple approach does not take into account the potential for organics to be lost during impact heating and pyrolysis. Impact models [164] showed that sizeable amounts of cometary matter can survive impacts into a terrestrial ocean at velocities ≤10 km/s. Under these conditions and on the time-scale of an impact, shock tube experiments [192] show that at least parts of simple organics, including short-chain aliphatic and aromatic compounds, can be expected to survive. We therefore regard the reported values here as the mass transported to the upper terrestrial atmosphere and representing an upper limit for the mass of volatile molecules arriving at the surface unaltered, given that many of these objects impact Earth at higher velocities [13].

Another hypothesis, the comet pond model [193], is based on the rare occasion of a soft entry of a comet at a low relative velocity with respect to Earth. A slow impact leads to a depression filled with organic-rich water after melting of the comet's ices [194]. With at least parts of cometary molecules originating in the cold environment of a prestellar core, rapid chemical reactions can be expected in the warm and liquid environment of the primitive Earth. Such a pond model may be appealing due to higher concentrations of various cometary organic molecules without the dilution associated to an impact in a terrestrial ocean. A shallow heterogeneous pond, containing sizeable concentrations of organics in contact with the minerals at the bottom of the pond and the interaction with solar UV, and weathering effects, may put in place the right conditions for the first metabolic systems to emerge. The downside of this model; however, is that it is based on a low probability scenario of high potential importance [194], while most of the organic mass may still have been delivered by more violent impacts.

Finally, the question of whether comet 67P/C-G can be considered representative of the cometary reservoir is of central concern when using the properties of this particular comet to extrapolate to some generic type of cometary material. Relative abundances are known to differ among comets and hence, change relative proportions of the volatiles delivered to Earth. During their journeys within the solar system, comets evolve under the combined effects of irradiation, erosion and specific thermal evolution, leading to the destabilization of volatile species and inhomogeneous subsurface compositions [65]. Comet 67P/C-G, as observed today, may therefore not exactly represent the pristine cometary material that could have been supplied to the early Earth. From the Kr and Xe isotopes measured within comet 67P/C-G, it is clear that the noble gas components are nucleosynthetically distinct from other solar system bodies [65, 71]; therefore arguing for a prestellar/protostellar origin of cometary ice with limited secondary processing. From the detailed analysis of one comet, it cannot be stated with absolute certainty that all comets will have the same composition as comet 67P/C-G. However, nucelosynthetically distinct signatures are not something that forms in isolation, and are most likely to originate from large cosmochemical reservoirs. Comets that formed within the same reservoir as comet 67P/C-G should therefore have, to a greater or lesser extent, a similar composition. Estimates of cometary accretion to Earth presented in this study can therefore be considered a basis for understanding the role comets play in delivering biologically relevant molecules to Earth. Additional missions to cometary bodies including the potential of a sample return will hopefully be able to confirm the representativeness of comet 67P/C-G and provide further weight to our initial estimates.

## 4.2. Prebiotic chemistry

Synthesis of prebiotic molecules potentially originated from abiotic organic molecules, radicals, and ions [195] in multiple regions and during several evolutionary phases of the solar system, including the Solar nebula [196] and/or the early Earth's surface [197]. Precursors of amino acids (the basis for proteins), lipids, and ribonucleotides (the building blocks of, e.g., adenosine triphosphate, ribonucleic acid (RNA), and deoxyribonucleic acid), can all be derived from (photo)chemistry of hydrogen cyanide, HCN, with hydrogen sulfide, $H_2S$, as a reductant [197]. The reductive homologation of HCN and some of its derivatives provides the $C_2$ and $C_3$ sugars, glycolaldehyde and glyceraldehyde, for later ribonucleotide assembly and formation of the precursor molecules of the amino acids glycine, alanine, serine, and threonine. As Patel, et al. [197] further elaborated, the formation of acetone is the basis for the precursors of the amino acids valine and leucine; while the first precursors of lipids could have been synthesized by phosphorylation of glycerol, an essential component of all cell membranes. Further copper-catalyzed reactions can then lead to the precursors of the amino acids proline, arginine, asparagine, glutamine, and glutamic acid.

In general, catalyzing properties of surface minerals probably played a key role in the emergence of protocells by i) supporting the formation of prebiotic reduced carbon- and nitrogen-bearing compounds, e.g., on metal-sulfide minerals [198], ii) protecting and stabilizing relevant biomolecules, e.g., ribose [199], (iii) promoting the chain elongation of bio-polymers [200], (iv) establishing structural micro-habitats acting as the earliest cell-like compartments, e.g., between mica sheets [201], and v) catalyzing the assembly of amphiphilic molecules into vesicles, resulting in the formation of proto-cell membranes [202]. In addition to protecting biomolecules from hydrolysis, UV light and thermal degradation [203], charged surfaces of clays could have catalyzed the formation of peptides [204] and RNA polymers (e.g., Ferris [205]).-Suitable natural conditions for the evolution of prebiotic molecules on early Earth, if achieved at some point, would most likely be reached in local and transient environments. For instance, $CH_4$, $H_2$, and metal sulfides could have been abundant close to black smoker vents (pH $< 3$, T $> 250$°C), providing reducing conditions for the formation of prebiotic molecules [203]. Likewise, the conditions at white smokers (pH $> 9$, T $< 90$°C) may have enabled formose reactions (self-condensation of formaldehydes into sugars) [203].

Interestingly, the two basic molecules, hydrogen cyanide (HCN) and hydrogen sulfide ($H_2S$), are both quite common in the volatile phase among comets [30] including 67P/C-G (cf. Table 4). The estimated masses delivered to Earth are in the range of $1.4 \cdot 10^{17}$ to $2.8 \cdot 10^{18}$ g and $1.4 \cdot 10^{18}$ to $2.8 \cdot 10^{19}$ g for HCN and $H_2S$, respectively. Both molecules are highlighted in blue in Figure 2. Furthermore, a subset of the intermediate species, namely ammonia [30] ($NH_3$), formaldehyde [30] ($H_2CO$), phosphorous oxide [81] (PO), acetylene [30, 61] ($C_2H_2$), glycolaldehyde [31] ($CH_2OHCHO$), acetone [36, 85] ($CH_3COCH_3$), and cyanoacetylene [27] ($HC_3N$), have been identified in comets (Table 1). Also, the amino acid glycine belongs to the organic (semi-)volatile inventory of comets [23, 81].

For a subset of these molecules, the relative abundances in 67P/C-G can be found in Table 4 and the upper limits of the mass delivered to Earth in Figure 2 (possibly including some isomers).

High-temperature reactions of carbonaceous meteoritic material with atmospheric nitrogen upon impact may have played an important role in the formation of hydrogen cyanide during the LHB or may originate in a geochemical scenario involving ferrocyanide evaporation and thermal metamorphosis due to geothermal activity, while phosphates may be the result of P-bearing minerals in iron-nickel meteorites dissolving in terrestrial surface water [197]. These molecules; however, could also have been delivered directly to the surface of the early Earth by comets.

## 4.3. Biosignatures

The molecules found in comets may not only be important for the emergence of life on Earth, but may also be relevant when searching for the presence of life elsewhere [206]. The number of detected exoplanets is increasing fast and includes numerous objects in the habitable zone around their host star [207]. This initiated the first studies on whether or not we can detect signs of life on such distant objects.

One of the approaches is to analyze the atmosphere of exoplanets for the presence of biologically relevant constituents. Biosignature gases are volatile molecules that form as a result of biological processes and then accumulate in the atmosphere, possibly in disequilibrium with the environment, where they could potentially be remotely detected. Seager, et al. [80] compiled a list of terrestrial biosignature gases including the corresponding major atmospheric sources. The origin of this list of molecules in the terrestrial atmosphere is quite diverse, while some are predominantly formed anthropogenically, others are produced by plants, but also from abiotical processes including the outgassing of the Earth's mantle and volcanism. The detection of gaseous molecules out of thermodynamical equilibrium, especially $O_2$ and $CH_4$, has been proposed to be a sign of life [208-210]. However, this is not without problems, as elaborated by Seager, et al. [80] for the example of methane, which can originate from methanogenesis by microbes and via abiotic chemistry in hydrothermal vents. Furthermore, while species such as $O_2$ and $CH_4$ may be major biosignature gases on Earth, Seager, et al. [80] emphasize that the situation may present itself differently on an exo-Earth.

The task becomes even more difficult when comparing the list of alleged biosignature gases with the volatile molecules detected in comets and in the ISM. Table 1 marks the subset of cometary species suggested to be potential biosignatures by Seager, et al. [80]. These molecules are produced through biological processes on Earth, but their presence in the atmosphere may be predominantly accounted for by abiotical processes. Cometary impacts may further alter a primordial exoplanetary atmosphere, if not overprinting it altogether, furnishing fresh material instead, as proposed by Kral, et al. [211] for the example of the TRAPPIST-1 system. The abundance of $CH_4$ varies among comets (less is known about $O_2$) and the cometary $O_2/CH_4$ ratio may be out of equilibrium for the conditions prevailing in the atmosphere of an exoplanet. A cometary contribution to a given exoplanet may therefore induce a local and temporal change in the atmospheric $O_2/CH_4$ and other species after impact.

Interestingly, several of the alleged biosignature volatiles in comets may be inherited from the earlier prestellar and protostellar phases. Fayolle, et al. [212] investigated the case of $CH_3Cl$, which is mainly produced by industrial and biological processes on Earth [213], whereas in comets, it most likely originates from the preceding phases of the Solar System's evolution as suggested by the detection of $CH_3Cl$ towards the low-mass protostar IRAS16293-2422 B (cf. section 3.2).

Some of the major biosignature molecules, such as $CH_4$ and $O_2$, are rather abundant in comets[30, 70], and other species such as the photochemically produced $O_3$ may even form through secondary reactions given the proper environment. The situation is similar for the sulfur-bearing molecules listed by Seager, et al. [80], which were all identified in comet 67P/C-G [67], except maybe dimethylsulfide, $(CH_3)_2S$, which could also be present in the form of its isomer ethanethiol, $CH_3CH_2SH$. In particular, also $CS_2$, which on Earth is predominantly of anthropogenic origin, is regularly observed in comets [30] (cf. Table 4).

Many species in the Earth's atmosphere can be found in comets, in particular the more abundant ones. As a consequence, the focus has to be put on the rare (ppt concentrations) and predominantly anthropogenically produced molecules, which renders the detection of life elsewhere extremely difficult, unless exocometary contributions to the atmosphere of an exoplanet can be ruled out for other reasons.

# 5. Conclusions

In this paper, we compared an updated inventory of volatile species found in and around comets, in particular 67P/C-G [124] and the ISM [62]. We investigated the origin of this material and how it links to the earlier prestellar and protostellar phases of formation of our Solar System [36]. We then estimated the mass of cometary material transported to the early Earth [12, 71], reviewed the relevance of cometary molecules for prebiotic chemistry [197], and discussed how the subset of cometary biosignature molecules may hinder our search for life elsewhere [80]. In summary:

- Comets contain a variety of volatile species including a sizeable fraction of organics [36]. In fact, about half of the detected molecules are COMs (organic molecules with at least six atoms including one or more carbon atoms). More than half of the cometary species, or at least one of their respective isomers, have been observed in the ISM [62].
- Comparing relative abundances of molecules in the ISM versus in comets, e.g., Hale-Bopp [27] and 67P/C-G [42], gives evidence for the formation in a similar environment and hence, through common chemical pathways. This is further supported by the presence of $O_2$ in the ice of comet 67P/C-G [70] and possibly 1P/Halley [127]. $O_2$ is thought to be of prestellar or protostellar origin [118, 135], in line with the measured D/H ratios in $D_2O$/HDO and HDO/$H_2O$ that are likely inherited from the earlier cold temperature phases of star formation [116].
- If cometary volatiles are largely preserved from the earlier evolutionary stages of our Solar System (cf. Drozdovskaya, et al. [42]), then they are initially formed through the astrochemical pathways operating in prestellar and protostellar environments in the cold-temperature regime. This includes neutral-neutral and ion-molecule reactions in the gas phase; and grain-surface hydrogenations and radical-radical associations. Some modification may occur during the time it takes to agglomerate the cometary body.
- Transport of material by comets to the early Earth may have been significant. The isotopes of the noble gas xenon indicate a $(22 \pm 5)\%$ contribution of cometary xenon to the terrestrial atmosphere [71]. Based on the relative abundance of the volatile species in the ice of comet 67P/C-G, we estimated the transport of different volatile species to the early Earth (cf. Marty, et al. [12]). The results suggest a minor water contribution of less than 1% to the Earth's oceans, whereas the contribution of biomolecules may be sizeable, i.e., up to the terrestrial biomass when destruction processes during the atmospheric entry and the impact at the surface are neglected. Also, the composition of comet 67P/C-G may not be representative of an average comet, but to date, it is the only one for which we have a comprehensive set of noble gas data and a detailed inventory of organic species.

- Several volatile molecules found in comets may be crucial ingredients for prebiotic chemistry. Patel, et al. [197] showed that a suite of amino acids can be derived by (photo)chemistry of HCN with $H_2S$ as a reductant. Both molecules may have a terrestrial origin, but are also commonly found in comets [30]. HCN and $H_2S$ may thus have been supplied by comets, together with a suite of relevant molecules, including intermediates in the reaction scheme proposed by Patel, et al. [197] up to the most basic amino acid glycine [23, 81]. Rare events, such as a soft impact on Earth, may have led to ponds of concentrated cometary organics [194] further facilitating the formation of even more complex organic molecules.
- (Exo)cometary impacts may significantly alter the atmospheres of exoplanets [211]. Comets contain numerous volatiles that have been put forward as potential biosignatures to be searched for in distant exoplanetary atmospheres [80]. Their sizeable abundance in the abiotic environment of comets further complicates the search for life elsewhere [212].
- On the other hand, if indeed comets played an important role in the emergence of life through the transport of pre-biotic molecules to the early Earth (and possibly Mars), similar processes may be occurring elsewhere.

Several of the molecules present in comets have not yet been identified in the ISM and vice versa. However, the strong link between the two suggests many more molecules to be commonly present. Especially, when taking into account the large number of currently unidentified lines in spectra of interstellar sources, e.g., obtained by powerful facilities such as ALMA . The detection of additional joint molecules is a matter of extending laboratory and theoretical spectroscopic studies.

# Acknowledgements

MR acknowledges the support of the State of Bern and the Swiss National Science Foundation (200021_165869, 200020_182418). DVD and MWB acknowledge the support from the European Research Council (PHOTONIS project, grant agreement No. 695618 to Bernard Marty). MND acknowledges the financial support of the SNSF Ambizione grant 180079, the Center for Space and Habitability (CSH) Fellowship and the IAU Gruber Foundation Fellowship. SFW acknowledges the financial support of the SNSF Eccellenza Professorial Fellowship PCEFP2_181150. All ROSINA data are the work of the international ROSINA team (scientists, engineers and technicians from Switzerland, France, Germany, Belgium and the US) over the past 25 years, which we herewith gratefully acknowledge. Rosetta is an ESA mission with contributions from its member states and NASA.